  \documentclass[final,5p,times,sort&compress]{elsarticle}

\usepackage{amssymb,amsmath}
\usepackage{lineno}
\usepackage[T1]{fontenc}
\usepackage[utf8]{inputenc}

\journal{Computational and Theoretical Chemistry}

\newcommand\equ[2]{\begin{align}\label{#1} #2 \end{align}}

\newcommand\figA[7]{
  \centering%\fbox{
  \includegraphics[width=#7,height=#6,
                   keepaspectratio=true,trim=#2 #3 #4 #5, clip=true]{#1}
}

\newcommand\ru           {{\b{r}_1}}
\newcommand\rd           {{\b{r}_2}}

\newcommand\rA           {{\b{r}_k}}
\newcommand\rR           {{\b{r}_{0}}}

\renewcommand\b[1]       {\mathbf{#1}}

\newcommand\ie           {\textit{i.e.}\ }

\newcommand\etc          {\textit{etc}\ }

\everymath{\displaystyle}
\newcommand\intext[1]    {$\textstyle #1$}

\newcommand\Ref[1]       {Eq.~(\ref{#1})}

\newcommand\SRef[1]      {\ref{#1}}
\newcommand\FRef[1]      {\ref{#1}}

\newcommand\expect[1]    {\left\langle #1 \right\rangle}
\newcommand{\bra}[2][]   {\mathinner{\langle #2|}_{#1}}
\newcommand{\ket}[2][]   {\mathinner{|#2\rangle}_{\hspace{-0.1em}#1}}

\newcommand\capALLchi{
Sectional drawings, for the molecules of
ethylene (top left),
butadiene (top right),
water (bottom left)
and naphthalene (bottom right),
along three orthogonal planes of the isocontours
    of (1) the response function corresponding
           to the reference point $\rR$ indicated by the black dot labelled "Q"
and of (2) the mathematical object defined by the right-hand side of
           equation \Ref{eq:chiApprox}, shown for the same reference point $\rR$.
The isocontour levels are in the range $\pm 0.002$ for the response function and static form factor of the ethylene and butadiene molecules, in the range $\pm 0.005$ for the response function and static form factor of the water molcule, in the range $\pm 0.0005$ for the response function of the naphthalene and $\pm 0.00025$ for the static form factor of the naphthalene. The isocontours are separated by $0.0005$ in the cases of the ethylene, butadiene and water molecule, by $0.0001$ for the response function of the naphthalene and $0.00005$ for its static form factor.
}

\newcommand\capALLhx{
Sectional drawings, for the molecules of
ethylene (top left),
butadiene (top right),
water (bottom left)
and naphthalene (bottom right),
along three orthogonal planes of the isocontours
    of (1) the exchange hole corresponding to
           the reference point $\rR$ indicated by the black dot labelled "Q"
and of (2) the square of the localized orbital which centroid is situated
           in this space domain.
The isocontours are in the range $\pm 0.03$ and separated by $0.005$ for the ethylene molecule, in the range $\pm 0.05$ and separated by $0.005$ for the butadiene, and in the range $\pm 0.005$ and separated by $0.0005$ for both the water and naphthalene molecule.
}

\newcommand\figALLchi{
\begin{figure*}[!htb]
\centering
\begin{minipage}[h]{.48\linewidth}
  \figA{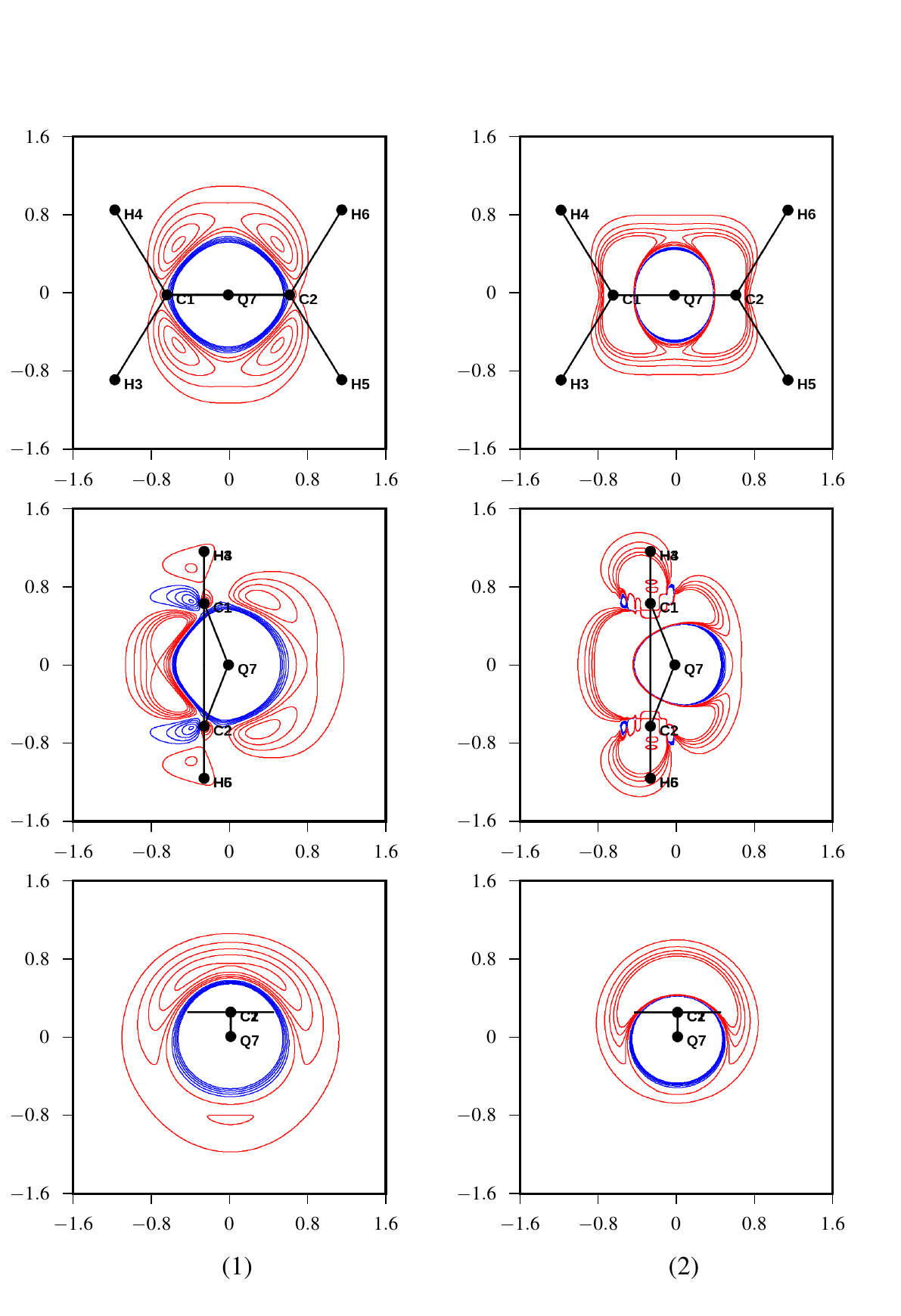}{0cm}{5mm}{0cm}{17mm}{20cm}{.9\linewidth}
\end{minipage}\qquad
\begin{minipage}[h]{.48\linewidth}
  \figA{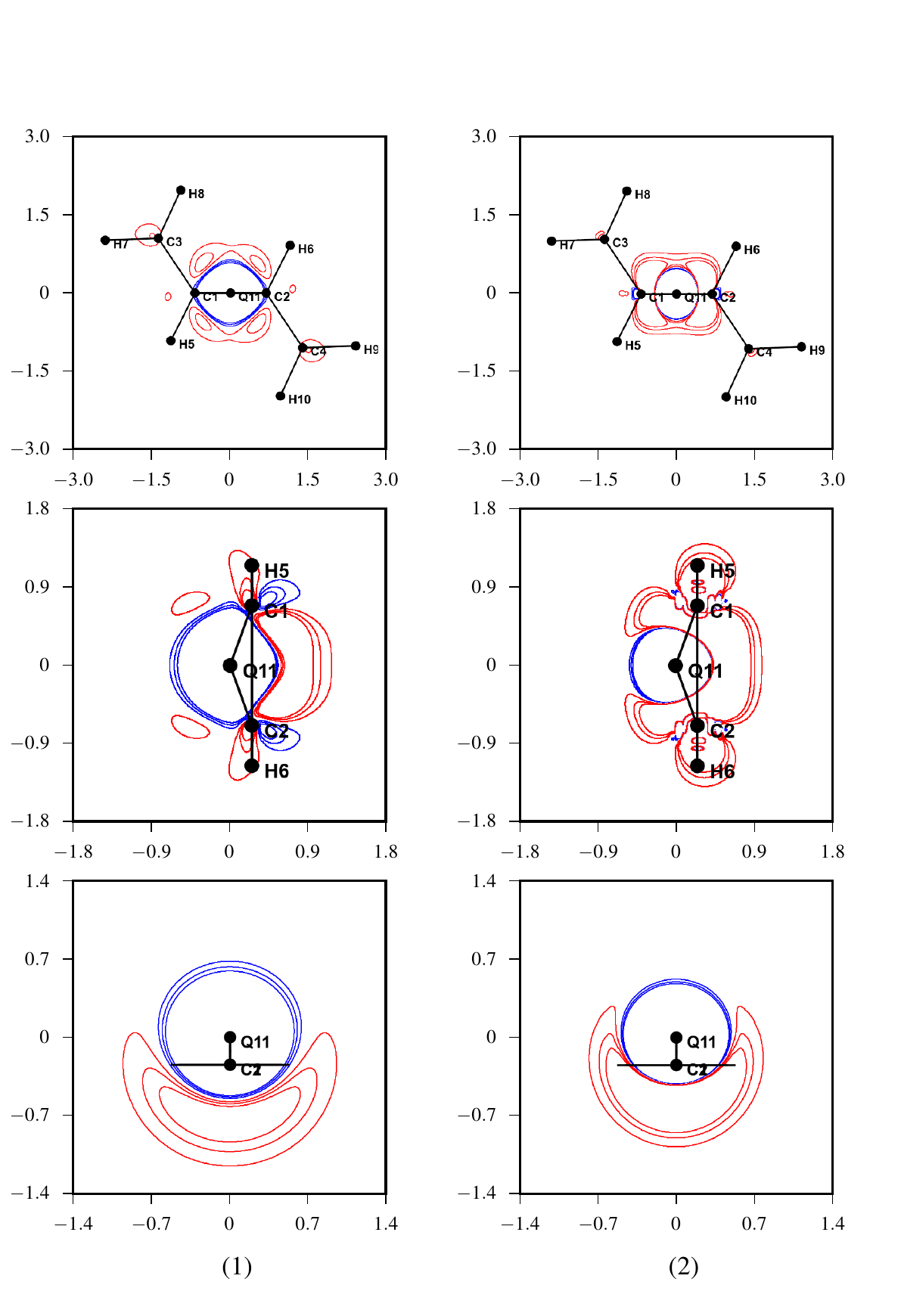}{0cm}{5mm}{0cm}{17mm}{20cm}{.9\linewidth}
\end{minipage}

\vspace{7mm}

\begin{minipage}[h]{.48\linewidth}
  \figA{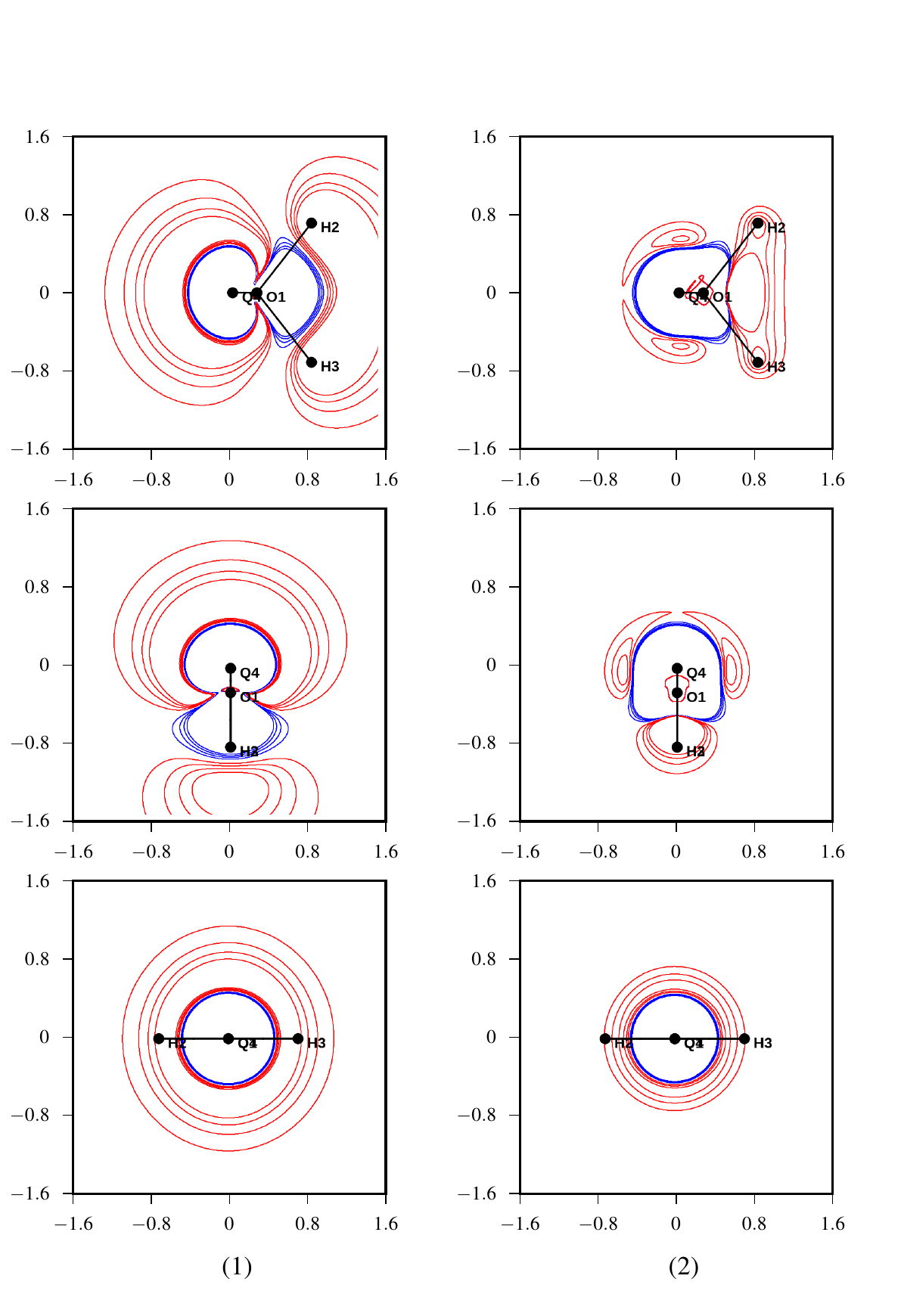}{0cm}{5mm}{0cm}{17mm}{20cm}{.9\linewidth}
\end{minipage}\qquad
\begin{minipage}[h]{.48\linewidth}
  \figA{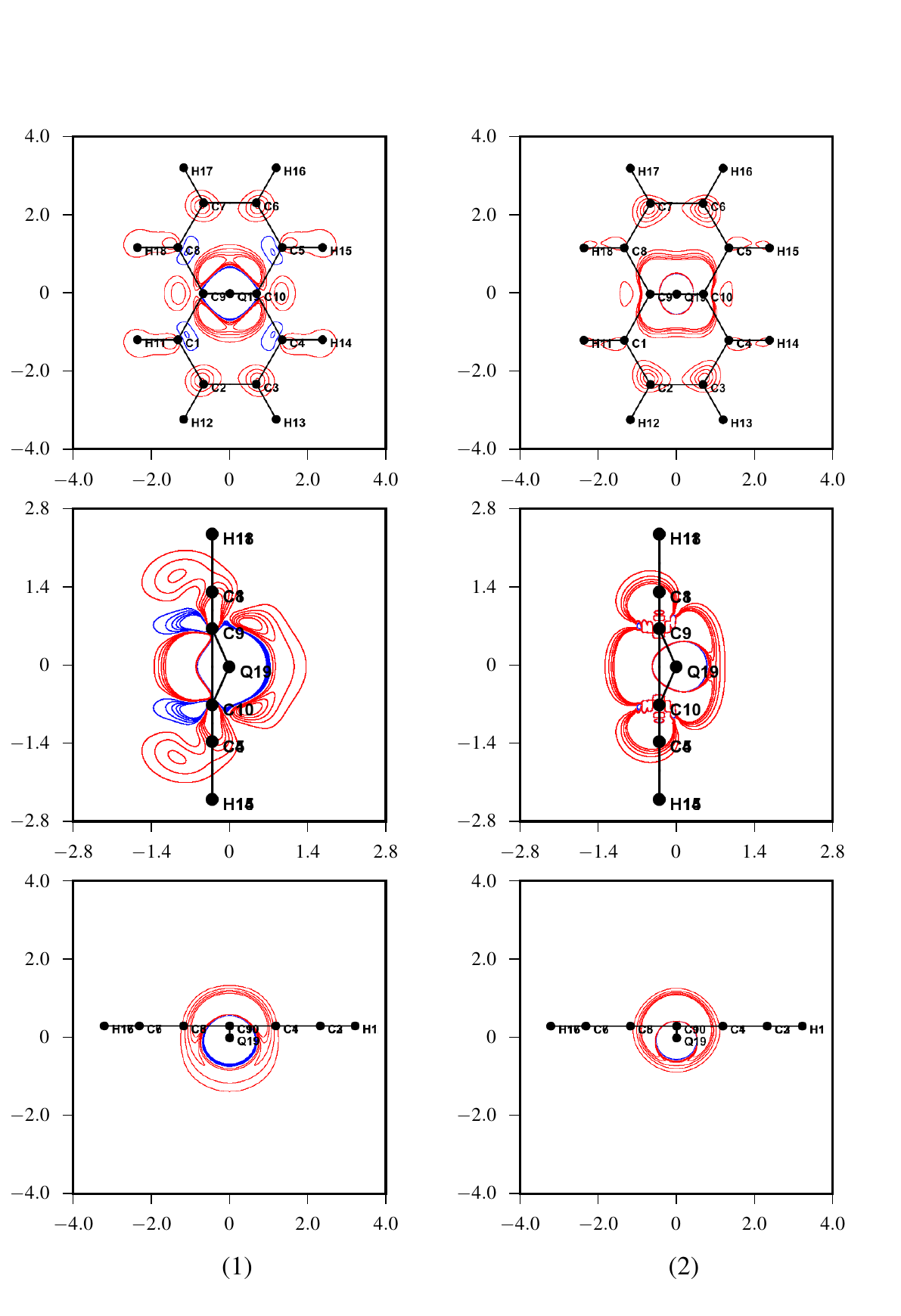}{0cm}{5mm}{0cm}{17mm}{20cm}{.9\linewidth}
\end{minipage}

\caption{\capALLchi}
\label{chiALL}
\end{figure*}
}

\newcommand\figALLhx{
\begin{figure*}[!htb]
\centering
\begin{minipage}[h]{.48\linewidth}
  \figA{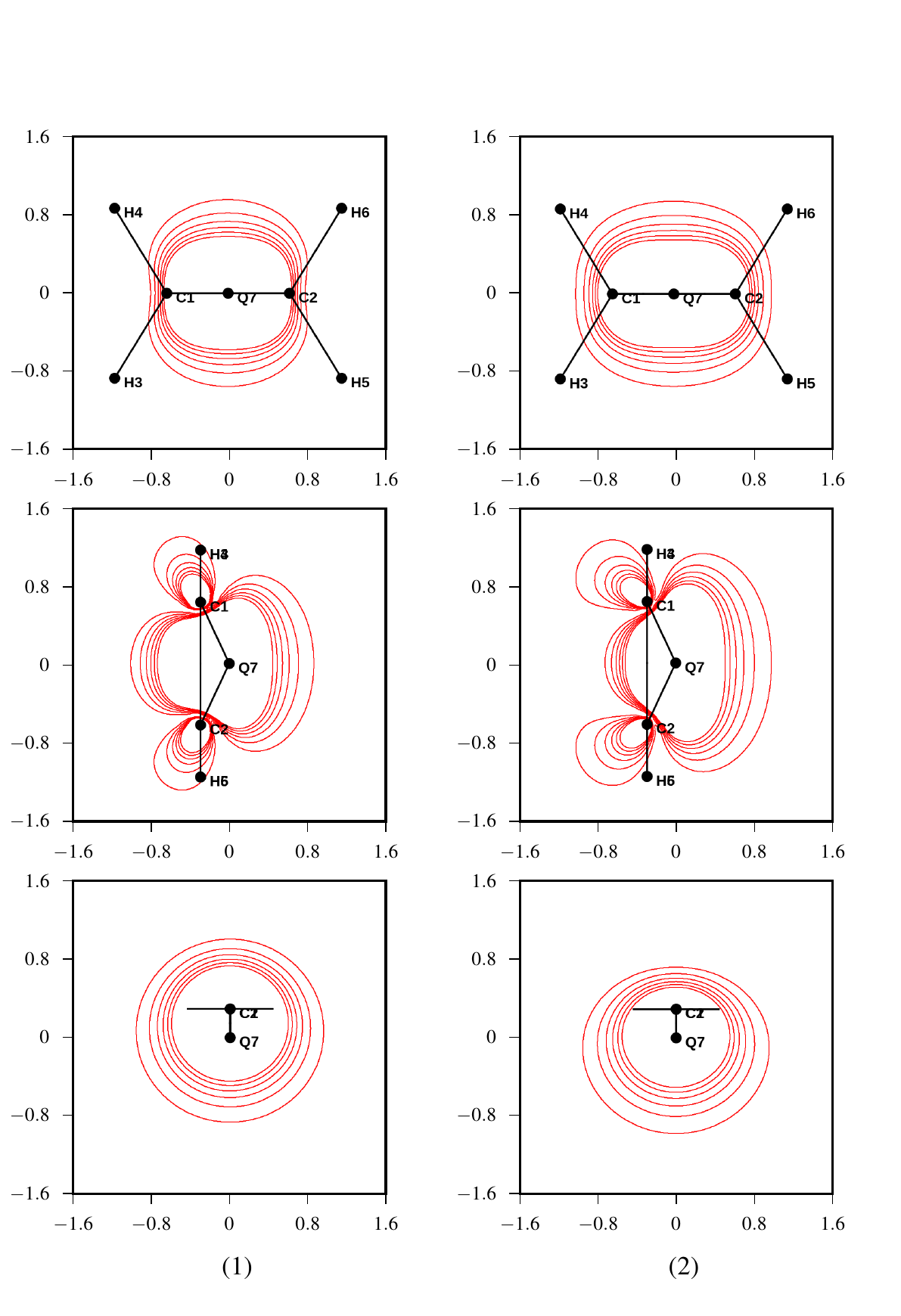}{0cm}{5mm}{0cm}{17mm}{20cm}{.9\linewidth}
\end{minipage}\qquad
\begin{minipage}[h]{.48\linewidth}
  \figA{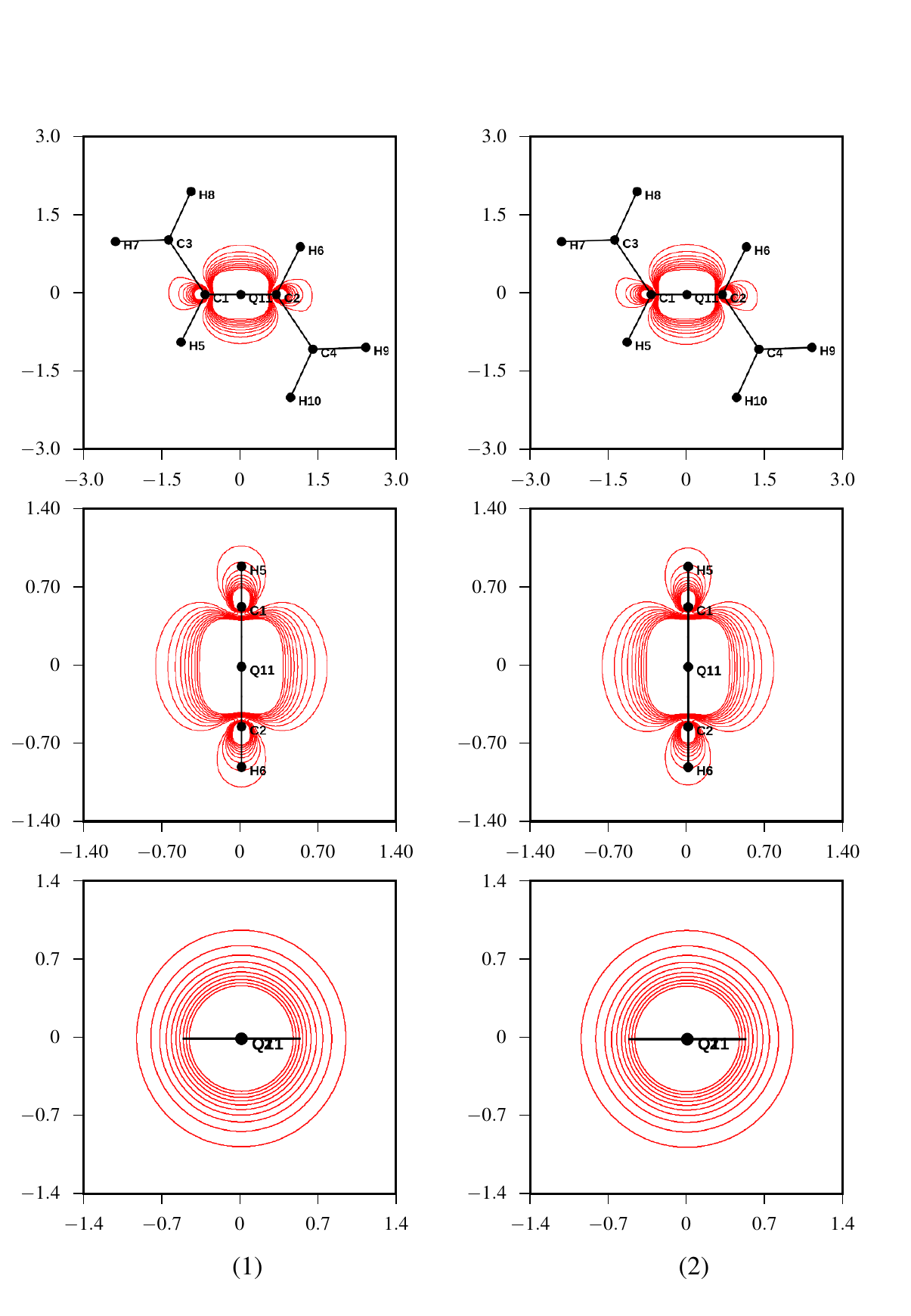}{0cm}{5mm}{0cm}{17mm}{20cm}{.9\linewidth}
\end{minipage}

\vspace{7mm}

\begin{minipage}[h]{.48\linewidth}
  \figA{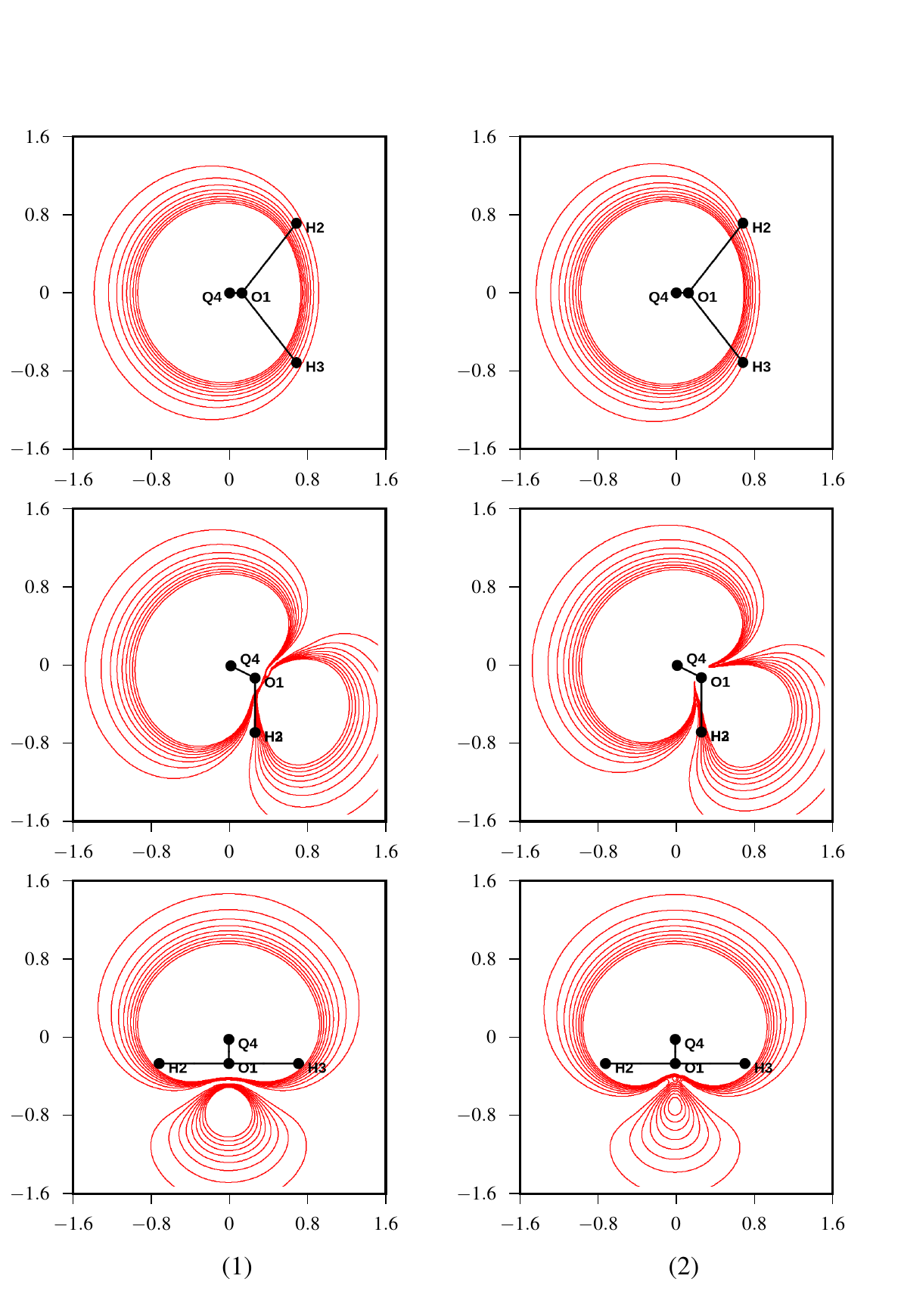}{0cm}{5mm}{0cm}{17mm}{20cm}{.9\linewidth}
\end{minipage}\qquad
\begin{minipage}[h]{.48\linewidth}
  \figA{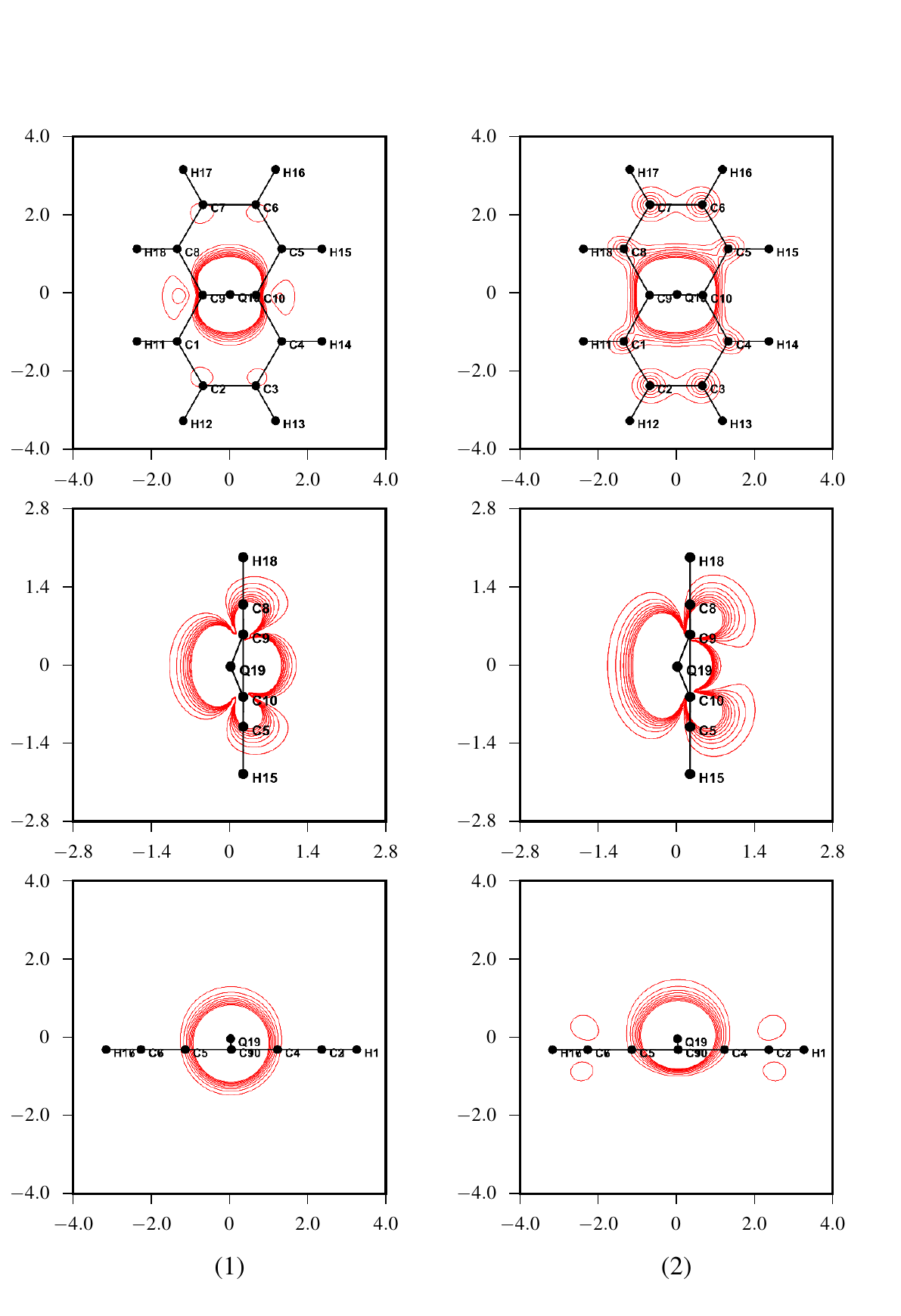}{0cm}{5mm}{0cm}{17mm}{20cm}{.9\linewidth}
\end{minipage}

\caption{\capALLhx}
\label{hxALL}
\end{figure*}
}

\begin{document}

\begin{frontmatter}



\title{Relationships between charge density response functions, \\ exchange holes and localized orbitals}


\author{Bastien Mussard\fnref{label1,label2,label3}}
\author{J\'anos G.~\'Angy\'an\corref{cor1}\fnref{label3,label4,label5}}
\ead{janos.angyan@univ-lorraine.fr}
\cortext[cor1]{Corresponding author}

\address[label1]{Institut du Calcul et de la Simulation, Université Pierre et Marie Curie, 4 place Jussieu, 75005 Paris, France}
\address[label2]{Laboratoire de Chimie Théorique,        Université Pierre et Marie Curie, 4 place Jussieu, 75005 Paris, France}
\address[label3]{IJB, CRM2, UMR 7036, Université de Lorraine, Vandoeuvre-l\`es-Nancy, France}
\address[label4]{IJB, CRM2, UMR 7036, CNRS, Vandoeuvre-l\`es-Nancy, France}
\address[label5]{Department of General and Inorganic Chemistry, Pannon University, Veszpr\'em, H-8201, Hungary} 

\begin{abstract}
 The charge density response function and the exchange hole are 
 closely related to each other via the fundamental 
 fluctuation-dissipation theorem of physics. A simple approximate 
 model of the static response function is visually compared on several
 examples in order to demonstrate this relationship.  
 This study is completed by illustrating the well-known  
 isomorphism between the exchange hole and the square of the dominant
 localized orbital lying in the space region of the reference 
 point of the exchange hole function. The implications of these 
 relationships for the interpretation of common
 chemical concepts, such as delocalization, are discussed. 
\end{abstract}

\begin{keyword}
  charge density response function \sep 
  Fermi-hole \sep 
  localized orbitals \sep 
  electron localization \sep 
  topological analysis 



\end{keyword}

\end{frontmatter}

\section{Introduction}
\label{sec:intro}

 To establish links between the wave function and chemical concepts 
 characterizing the structure and reactivity of atoms, molecules,
 solids and interfaces has always been a considerable challenge for
 computational and theoretical chemistry. Although this is a 
 subject which has been in the focus of scientists since the
 very beginning of quantum chemistry, the development of Conceptual
 Quantum Chemistry tools remains even in our days a dynamically 
 developing domain, producing new ideas and leading to a deeper
 understanding of old concepts. This research activity betrays 
 a natural need for having a sound basis of quantitative or 
 semi-quantitative characterization of chemical objects and their
 interaction in the framework of concepts, which are most often 
 deeply rooted in chemical thinking. Technically it means that one
 has to extract specific parameters from many-electron wave 
 functions of high complexity, which are able to quantify or 
 illustrate concepts. 
    
 If we are to identify some trends in the evolution 
 of conceptual tools over the decades,
% since the dawn of 
% theoretical chemistry and theoretical material sciences
 one of the most remarkable feature is that the definition of
 extracted quantities become more and more independent of any 
 specific approximation scheme and are rooted more and more 
 in true observables.  On the one hand, at the beginning, quantities like
 bond orders and atomic populations were constructed 
 directly from intermediate quantities of popular quantum
 chemical approximations (molecular orbitals, populations analysis 
 in a given basis set, valence bond configurations, \dots)
 which could not be interpreted outside the scope of a given
 computational scheme. On the other hand, in the course of the last 
 20 years, 
 researchers made an effort to define generally applicable
 conceptual measures and indices, which are valid independently 
 of any quantum chemical approximation method. Emblematic examples 
 of such approximation-independent concepts are  
 atomic multipoles defined on the basis of the topological 
 partitioning of the electron density \cite{Bader:book}, 
 delocalization indices between topologically defined 
 atoms \cite{Bader:75,Cioslowski:91c,Angyan:94b,Bader:96,Fradera:99} 
 or, more recently, the total position spread 
 (TPS) \cite{Brea:13}, based on the localization tensor of 
 Resta \cite{Resta:06b}. 
 Various measures of electron localization, like the electron 
 localization function (ELF) \cite{Becke:90,Savin:91,Silvi:94}, 
 and other  analogous functions  of  
 the three-dimensional space \cite{Schmider:02,Kohout:04,Ayers:05} 
 are also defined in a universal 
 manner, without referring to specific notions which would be
 valid only in the realm of a particular approximate electronic 
 structure method. Such a philosophy has not only the advantage 
 of making possible a judicious comparison of results obtained 
 from different electronic structure methods (e.g.\ atomic charges
 obtained from Gaussian and plane-wave basis set calculations), 
 but opens the way to the comparison of theoretical values directly with the
 experiment. For instance, atomic charges and Laplacians derived from the 
 QTAIM (Quantum Theory of Atoms In Molecules) of Bader can be 
 derived not only from 
 computed, but also from experimental electron densities obtained 
 from high-resolution X-ray dif\-fraction data 
 \cite{Kuntzinger:98,Volkov:00a,Gatti:05,Garcia:07}.

 Inspired by the works of Parr and his coworkers 
 \cite{Parr:94:book}, a whole family 
 of functions have been defined in the framework of the 
 Conceptual Density Functional Theory, leading to a formally 
 rigorous interpretation of the chemical potential of an electronic 
 system as the functional derivative of the electronic energy with 
 respect to the number of electrons, and of the linear charge density 
 response function as the functional derivative of the
 electronic energy with respect to the external potential 
 \cite{Senet:97}. One can define 
 quantities like hardness, softness, Fukui function, \dots \textit{via} 
 various higher order and mixed derivatives \cite{Senet:96}.
 It has to be stressed that these definitions are fully general and 
 valid not only in DFT but, in 
 principle, at any level of modern electronic structure theory.   
 Among all these quantities, we are going to be
 concerned mainly by the softness, more precisely by the
 softness kernel.

 The relevance of the linear charge density response function,
 which is called by some authors \textit{linear response kernel} 
 and which is equal to the negative of the softness kernel, for the 
 characterization of localization/delocalization properties of an 
 electronic system has been recognized in the past by several 
 authors. The remarkable correlation between atom-atom softnesses 
 \ie atomic partition of the linear response functions and atom-atom 
 delocalization indices have been pointed out through the
 example of Y-conjugated compounds \cite{Angyan:00b,Angyan:11d}. 
 The properties of the linear density-density response and its 
 eigenvalue decomposition have been used by Savin and his 
 coworkers to analyze the strong density-dependence of the Kohn-Sham 
 potential in density functional theory \cite{Savin:03} and the 
 radial density-density response function has been analyzed for a 
 series of atoms along the adiabatic connection 
 path~\cite{Savin:01}. 
  
 More recently, in a series of papers~\cite{Sablon:10b,Sablon:10c,Sablon:12,Boisdenghien:13,Fias:13,Boisdenghien:14,Geerlings:14}, 
 Geerlings and his coworkers have analyzed the distribution
 of the charge density response in the space  for atoms and
 molecules. One possibility to study this quantity
 is to calculate the atom-condensed linear response parameters, 
 corresponding to the charge-flow polarizabilities in the general 
 distributed polarizability theory of 
 Stone~\cite{LeSueur:93, LeSueur:94}. Distributed polarizabilities can
 be implemented in the context of various definitions of atoms in 
 molecules (AIM). The possibilities cover a wide range of 
 definitions of AIM, going from the 
 partition of the basis functions 
 to the Hirshfeld-like fuzzy  \cite{Mayer:04b}
 or to the QTAIM \cite{Bader:book} discontinuous partitioning of the space. 
 QTAIM charge-flow polarizabilities 
 have been introduced first in \cite{Angyan:94c}. 
  
 The full non-expanded density-density response kernel, 
 calculated in the uncoupled perturbed Hartree-Fock theory, 
 has been plotted in Ref.~\cite{Fias:13} 
 for a few planar metallic and simple organic systems in view of  
 characterizing their aromaticity. The connection of the response
 kernel to the delocalization and to induction and resonance 
 effects have been discussed in Refs.~\cite{Sablon:10c,Sablon:10b}.
 
 In the present contribution, we discuss not only 
 a few more examples to illustrate the behavior 
 of the linear density-density 
 response kernel for molecular systems as plots of the 
 charge deformation due to a perturbation in a fixed point in the 
 space, but we attempt to point out and to demonstrate the possible 
 relationships between the response kernel and the exchange (or
 Fermi) hole function. Furthermore, the well-known connection 
 between the exchange hole and localized orbitals is also 
 illustrated.  

%%%%%%%%%%%%%%%%%%%%%%%%%%%%%%%%%%%%%%%%%%%
\section{Basic relationships}
\label{sec:basics}
%%%%%%%%%%%%%%%%%%%%%%%%%%%%%%%%%%%%%%%%%%%%

%%%%%%%%%%%%%%%%%%%%%%%%%%%%%%%%%%%%%%%%%%%%%%%%%%%%%%%%%%%%%%%%%%%%
\subsection{Exact response and exchange-correlation hole functions}
%%%%%%%%%%%%%%%%%%%%%%%%%%%%%%%%%%%%%%%%%%%%%%%%%%%%%%%%%%%%%%%%%%%%

 Using the definition of the two-particle (pair) density operator
 \intext{\hat{n}_2(\ru,\rd)} in terms of the one-particle 
 density operator \intext{\hat{n}(\ru)}
and of $\delta(\ru,\rd)$, the Dirac delta function 
%(see Eq.~(\ref{eq:delta})):
defined later:
 \equ{}{
 \hat{n}_2(\ru,\rd)&=
 \hat{n}(\ru)\hat{n}(\rd)-
 \delta(\ru,\rd)\hat{n}(\ru),}
 and the definition of the ex\-change-correlation hole in terms of the
 expectation values \intext{n(\ru)} and \intext{n_2(\ru,\rd)}:
 \equ{}{
 h_{xc}(\ru,\rd)&=
 \frac{n_2(\ru,\rd)}{n(\ru)}-n(\rd)
,} 
 one can deduce the following relationship: %  by inserting the 
% expectation values \intext{n(\ru)} and \intext{n_2(\ru,\rd)},:

\figALLchi

\equ{eq:trouXCmanip}{
n(\ru)h_{xc}(\ru,\rd)&=n_2(\ru,\rd)-n(\ru)n(\rd)
\nonumber\\ &=\expect{\hat{n}_2(\ru,\rd)}
             -\expect{\hat{n}(\ru)}\expect{\hat{n}(\rd)}
\nonumber\\ &=\expect{\hat{n}(\ru)\hat{n}(\rd)}
             -\delta(\ru,\rd)\expect{\hat{n}(\ru)}
             -\expect{\hat{n}(\ru)}\expect{\hat{n}(\rd)}
\nonumber\\ &=\expect{\delta\hat{n}(\ru)\delta\hat{n}(\rd)}
             -\delta(\ru,\rd)\expect{\hat{n}(\ru)}
,}
 where 
 \intext{\langle\hat{O}\rangle=\bra{\Psi_0}\hat{O}\ket{\Psi_0}} 
 denotes a ground state expectation value, 
 and we use the decomposition of the density operator 
 as a fluctuation $\delta\hat{n}(\ru)$ around its mean: 
 $\hat{n}(\ru)=n(\ru)+\delta\hat{n}(\ru)$.
 The Dirac delta function $\delta(\ru,\rd)=\delta(\rd-\ru)$ is 
 usually defined in the most general terms
 via its sifting property \cite{Boykin:03}
 
 \equ{eq:Diracsifting}{
 \int d\ru\, f(\rd) \, \delta(\rd-\ru) = f(\ru). }

 There are numerous options to represent $\delta(\rd-\ru) $ 
 by the limiting value  of a series of functions, like in Eq.~(\ref{eq:delta}).
 
  \Ref{eq:trouXCmanip} leads to:

\equ{eq:trouXCmanip2}{
 \expect{\delta\hat{n}(\ru)\delta\hat{n}(\rd)}
=\delta(\ru,\rd)n(\ru)+n(\ru)h_{xc}(\ru,\rd)
.}

 The right-hand side of the above equation is usually referred 
 to as the static form factor (cf.~Ref.~\cite{Angyan:07a}). As we can see,
 Eq.~(\ref{eq:trouXCmanip2}) establishes a relationship between the 
 density-weighted
 hole function and the \textit{fluctuations} of the charge density, 
 fluctuations that are themselves related to the response function by the 
 virtue of the (zero-temperature) flu\-ctu\-ation-dissipation theorem. 
 In the following we derive an approximate relationship between 
 \Ref{eq:trouXCmanip2} and the \textit{static} response function. 

 We  proceed by the technique of effective denominators,
 usually related to the name of Unsöld~\cite{Unsold:27}
 and which has been later generalized by others,
 e.g.~\cite{Linder:80,Malinowski:81,Berger:12}. 
 The aim of this technique is to replace the state-specific energy
 denominator in the sum-over-states expression of the response 
 function by a constant and to transform the sum over the excited 
 states in the numerator to a ground state expectation value by invoking
 the resolution of identity.

 Using the shorthand notation for the transition density
 associated with the $\alpha$ excited state
 $n_{\alpha}(\ru) = \bra{\Psi_0}\hat{n}(\ru)\ket{\Psi_\alpha}$
 and with the excitation energy $\omega_\alpha$, 
 the \textit{static} response function can be written as:

\equ{}{
\chi(\ru,\rd;0)          & =  2                                \sum_{\alpha\neq 0} \frac{n_{\alpha}  (\ru) n_{\alpha}  (\rd)}{\omega_{\alpha}}
\nonumber\\\nonumber     & =       \frac{2}{\omega(\ru,\rd)}   \sum_{\alpha\neq 0}       n_{\alpha}  (\ru) n_{\alpha}  (\rd)
\\\label{eq:OmegaDefPROV}& =       \frac{2}{\omega(\ru,\rd)}   \expect{\delta\hat{n}(\ru)\delta\hat{n}(\rd)}
\\\label{eq:chiStat}     & \approx \frac{2}{\overline{\omega}} \expect{\delta\hat{n}(\ru)\delta\hat{n}(\rd)}
,}  
 where, disregarded the hypothesis of working with real wave 
 functions, only the last step involves an approximation. While 
 the position-dependent effective denominator function 
 $\omega(\ru,\rd)$ in line (\ref{eq:OmegaDefPROV}) is able to 
 maintain the equality, its reduction to a position-independent constant 
 $\overline{\omega}$ necessarily leads to an approximate expression of 
 the response function.
 We have used similar approximations in earlier works
 \cite{Angyan:07a,Angyan:11d}.
 Comparing equations \Ref{eq:trouXCmanip2} and \Ref{eq:chiStat}
 allows us to write:

\equ{eq:chiApprox}{
 \frac{\overline{\omega}}{2}\chi(\ru,\rd;0    )
=\delta(\ru,\rd)  n(\ru)+n(\ru)h_{xc}(\ru,\rd)
,} 
 which is the desired (approximate) relationship between the 
 static response function and the right-hand side of 
 \Ref{eq:trouXCmanip2}, the static form factor. 
 Note that the two contributions to the static form factor,
 corresponding to the density multiplied, on the one hand, by 
 the Dirac delta function and on the other hand, by the xc-hole function, 
 ensure the correct charge conservation sum-rule. Indeed,
 the response function on the left-hand side of Eq.~(\ref{eq:chiApprox}) 
 integrates to zero, while, on the right hand side,
 the Dirac delta function integrates to plus one
 and the exchange-correlation hole function to minus one.

 In the following we are going to simplify the problem and instead of 
 working with the exact response function and the exchange-correlation 
 hole function, we are going to consider the noninteracting (or bare)
 response function and the exchange hole function, which arises 
 in an independent particle model.

 %%%%%%%%%%%%%%%%%%%%%%%%%%%%%%%%%%%%%%%%%%%%%%%%%%%%%%%%%%%%%%%%%%
 \subsection{Noninteracting response and exchange hole functions}
 %%%%%%%%%%%%%%%%%%%%%%%%%%%%%%%%%%%%%%%%%%%%%%%%%%%%%%%%%%%%%%%%%
 In this subsection we are going to apply the effective 
 denominator technique to the  zero-frequency bare or noninteracting 
 response function:

\equ{eq:chiunscreened}{
{\chi_0}(\ru,\rd) = 2
\sum_{i}^\text{occ}\sum_{a}^\text{virt}
\frac{\phi_a^\ast(\rd) 
      \phi_i(\rd)
      \phi_i^\ast(\ru)
      \phi_a(\ru)}{\epsilon_a-\epsilon_i}
,}
 where $\phi_i,\epsilon_i$ ($\phi_a,\epsilon_a$)
 are occupied (virtual) orbitals and orbital energies.
 Using a position-dependent effective denominator, $\omega(\ru,\rd)$,
 this yields:
      
\equ{eq:chiEED}{
\omega(\ru,\rd)\,{\chi_0}(\ru,\rd)=
 2\sum_{i}^\text{occ}\sum_{a}^\text{virt}
     {\phi_a^\ast(\rd) 
      \phi_i(\rd)
      \phi_i^\ast(\ru)
      \phi_a(\ru)}
.}

 The summation over the virtual orbitals is transformed into
 a sum over all the orbitals minus a sum over occupied orbitals:

\equ{eq:chisum}{
\omega(\ru,\rd)\, {\chi_0}(\ru,\rd)
&= 2\sum_{i}^\text{occ}\sum_{p}^\text{all}
     {\phi_p^\ast(\rd) 
      \phi_i(\rd)
      \phi_i^\ast(\ru)
      \phi_p(\ru)}
\nonumber\\&\quad
 - 2\sum_{i}^\text{occ}\sum_{j}^\text{occ}
     {\phi_j^\ast(\rd) 
      \phi_i(\rd)
      \phi_i^\ast(\ru)
      \phi_j(\ru)}
,}
 where $\phi_j$ is an occupied orbital and $\phi_p$ designates 
 an arbitrary (virtual or occupied) orbital.
 In this last expression, we recognize
 the usual definition
 of the exchange hole ${h_x(\ru,\rd)}$ for a single determinant 
 wave function:

\equ{eq:hx}{
{n(\ru)}\,h_x(\ru,\rd)   =-2 \sum_{ij}
                  \phi_i^\ast(\ru)\phi_j(\ru)\phi_j^\ast(\rd)\phi_i(\rd)
,}
 and
 the resolution-of-identity
 expression of the Dirac delta function:

\equ{eq:delta}{
\delta(\ru,\rd)=\sum_{p} {\phi_p^\ast(\rd) \phi_p(\ru)}.}

 Note that the above representation of the Dirac delta function 
 holds strictly only in the case of a complete orbital basis. 
 We obtain: 

\equ{eq:chi}{
\omega(\ru,\rd)\,{\chi_0}(\ru,\rd)&= 
  2{\delta(\ru,\rd)\sum_{i} \phi_i(\rd) \phi_i^\ast(\ru)}
 +n(\ru)h_x(\ru,\rd)
,}
 where we can use the Dirac delta function to see the
 expression of the charge density $n(\ru)$ emerge in the first term: 

\equ{eq:chifinql}{
\omega(\ru,\rd)\,{\chi_0}(\ru,\rd)=
  {\delta(\ru,\rd)n(\ru)}
 +n(\ru)h_x(\ru,\rd)
.}

 As seen before, the simplest approximation to this potentially exact relation is 
 to replace $\omega(\ru,\rd)$ by a position-in\-de\-pen\-dent constant,
 $\overline{\omega}$:

\equ{eq:chifinql2}{
\overline{\omega}\,{\chi_0}(\ru,\rd)\approx
  {\delta(\ru,\rd)n(\ru)}
 +n(\ru)h_x(\ru,\rd)
.}

 Compared to Eq.~(\ref{eq:chiApprox}),
 which relates the exact interacting static response function 
 to the exchange-correlation hole,
 equation~(\ref{eq:chifinql2}) involves the noninteracting 
 static response function and the exchange hole.
 
 Our average energy denominator approximation is expected to fail 
 in reproducing  
 the nodal structure of the res\-ponse function in all its details, but 
 the major features are expected to be preserved. In the present article
 all the numerical examples  will be limited to the noninteracting (bare) 
 response function and to the corresponding exchange hole. 

%%%%%%%%%%%%%%%%%%%%%%%%%%%%%%%%%%%%%%%%%%%%%%%%%%
\subsection{Exchange-hole and localized orbitals}
%%%%%%%%%%%%%%%%%%%%%%%%%%%%%%%%%%%%%%%%%%%%%%%%%%%

 The idea that localized orbitals are closely related to the 
 exchange hole has probably first appeared in the work of 
 Luken \cite{Luken:82a,Luken:90} at the beginning of the eighties
 and it has been later reiterated by Tschinke and Ziegler 
 \cite{Tschinke:89} as well as by others~\cite{Buijse:02}. 
 The central idea is that the definition of the exchange hole 
 for a single determinant wave function of \Ref{eq:hx}
 can be considerably simplified if it were possible to use
 strictly localized orbitals. Indeed, strictly localized orbitals 
 are such that the "differential overlap" 
 $\phi^\ast_i(\ru)\phi_j(\ru)$ is (almost) zero for $i\neq j$ 
 and it is $\vert \phi_i(\ru)\vert^2$ for $i= j$. 
 Therefore we can write the hole function as:

\equ{eq:Hxloc}{
h_x(\ru,\rd)=-2\sum_i 
              \frac{ \left| \phi_i(\ru) \right|^2}{n(\ru)}
              \left| \phi_i(\rd) \right|^2
.}

 Let us consider a partition of the physical 
 space to approximately disjoint regions 
 $\Omega_i$~\cite{Angyan:07a}, each belonging to a 
 strictly localized orbital $\phi_i$. In every region only one
 localized orbital contributes to the   electron density, \ie
 \intext{ \left\vert \phi_i(\ru) \right\vert^2\approx n(\ru)}. 
 Using the window function $\Theta_i(\ru)$ (which is equal 
 to unity if $\ru$ is in $\Omega_i$ and zero otherwise) it is 
 possible to write that:

\equ{eq:Hxloc2}{
h_x(\ru,\rd)=-2\sum_i \Theta_i(\ru) \left| \phi_i(\rd) \right|^2 
,}
 \ie that the exchange hole associated with a reference point 
 $\ru$ is approximated as the negative of the square of the 
 localized orbital $k$ which is situated in the space domain 
 of this point and satisfies $\Theta_k(\ru)=1$, that is to say:

\equ{eq:Hxloc3}{
h_x(\ru,\rd)=- 2 \left| \phi_k(\rd) \right|^2 \qquad \text{for 
  $\ru, \rd \in \Omega_k$}.
}

 In Section \SRef{sec:comparisons}, it is shown in several 
 illustrations how far this reasoning holds in different systems
 characterized by different degree of localization of their 
 electrons.

\figALLhx

%%%%%%%%%%%%%%%%%%%%%%%%%%%%%%%%%%%%%%%%%%%%%%%%%%%%%%%%%%%%%%%%%%%%%%%%%
\section{Picturial comparison of response functions,\\ exchange holes and localized orbitals}
\label{sec:comparisons}
%%%%%%%%%%%%%%%%%%%%%%%%%%%%%%%%%%%%%%%%%%%%%%%%%%%%%%%%%%%%%%%%%%%%%%%%%

 %%%%%%%%%%%%%%%%%%%%%%%%%%%%%%%%%%
 \subsection{Computational details}
 %%%%%%%%%%%%%%%%%%%%%%%%%%%%%%%%%%
 A software has been written to calculate 
 the two-variable functions $\chi_0(\rA,\rR)$, $\delta( \rA,\rR)$ and 
 $h_{x}(\rA,\rR)$ appearing in  equations (\ref{eq:chiunscreened}), 
 (\ref{eq:delta}) and (\ref{eq:hx}) on regular grids $\{ \rA \}$,
 for a fixed reference point $\rR $, 
 permitting us to generate
 visual representations of both sides of \Ref{eq:chiApprox} and 
 \Ref{eq:Hxloc3}. The program takes as input the orbitals 
 and the total electronic density of a system in the {\tt CUBE} 
 file format using the {\tt MOLPRO} program \cite{Werner:11m}. 
 The calculations have been done at the Kohn-Sham level using 
 the LDA functional and the aug-cc-aVTZ basis set.
 The sum rules \intext{\int d \rA \; \chi_0  (\rA,\rR) =  0},
           \intext{\int d \rA \; h_{x} (\rA,\rR) = -1}         
       and    
       \intext{\int d \rA \; \delta(\rA,\rR) =  1}
 have been checked to verify the suitability of the grid by simple 
 summation over the grid points. 
 The Dirac delta function has been represented 
 by a spherical Gaussian model:
 \intext{\delta( \rA,\rR)\approx \frac{w}{2\pi} 
 e^{\left(-\frac{1}{2} w^2 |\rA-\rR|^2\right)}}.
 The half-width parameter $w=2.5$ has been
 found on our grids to produce a normalization 
 integral closest to $1$ for all studied systems.
 Localized orbitals were generated using the Foster-Boys localization 
 criterium \cite{Foster:60a,Boys:66}.

%%%%%%%%%%%%%%%%%%%%%%%%%%%%%%%%%%%%%%%%%%%%%%%%%%%%%
\subsection{Response function and exchange hole}
%%%%%%%%%%%%%%%%%%%%%%%%%%%%%%%%%%%%%%%%%%%%%%%%%%%%%
 In order to visualize the relationship between the response 
 function and the hole function, we present contour plots of 
 the noninteracting response function,
\begin{align}
\chi_0(\rA,\rR;0)
\quad \text{and} \quad 
\delta(\rA,\rR)n(\rA) + n(\rA)h_x(\rA,\rR)
\nonumber
,
\end{align}
 in  cross sections of the molecule lying in the $xy$, $yz$ and 
 $zx$ planes. 

 Figure \FRef{chiALL} shows the correspondance between the 
 static response function $\chi(\rA,\rR;0)$
 and the static form factor, \ie the function appearing
 on the right-hand side of \Ref{eq:chiApprox}. We have used as 
 example the ethylene, butadiene, water  and naphthalene molecules.
 The reference points $\rR$ are labelled with a "Q" on the Figures.
 The quality of the ressemblance between the static response function
 and the static form factor does not depend on the choice of the 
 reference points, but these are placed on symmetry planes or axis for 
 each molecule and are chosen such that they coincide with 
 a grid point.
 Due to the relatively rough grids, an arbitrarily placed
 reference point would have destroyed the symmetry of the 
 resulting figure.
 For the ethylene and butadiene molecule the 
 reference point is near the C-C single bond, outside of the plane 
 of the molecule. In the case of the water molecule, the reference 
 point is on the bisector of the H-O-H angle. The reference point 
 for the naphthalene molecule is near the central C-C bond, above
 the molecular plane. 
 
 Overall, the correspondance between the two functions is of 
 rather acceptable visual quality, in spite of some discrepancies,
 which can be attributed to the relative crude model of the
 response function, supposing a simple, 
 position independent, proportionality with the
 form factor. The incompleteness of the model can be conjectured
 from the fact that linear response functions possess a richer 
 topological structure, as compared to the form factor based model.
 The mathematical structure of the form factor makes clear that the
 dominant feature in both functions should be a positive peak centered 
 on the reference point $\rR$  and an essentially negative region 
 corresponding to the exchange hole.

\begin{figure*}[!htb]
\centering
\begin{minipage}[h]{.48\linewidth}
  \figA{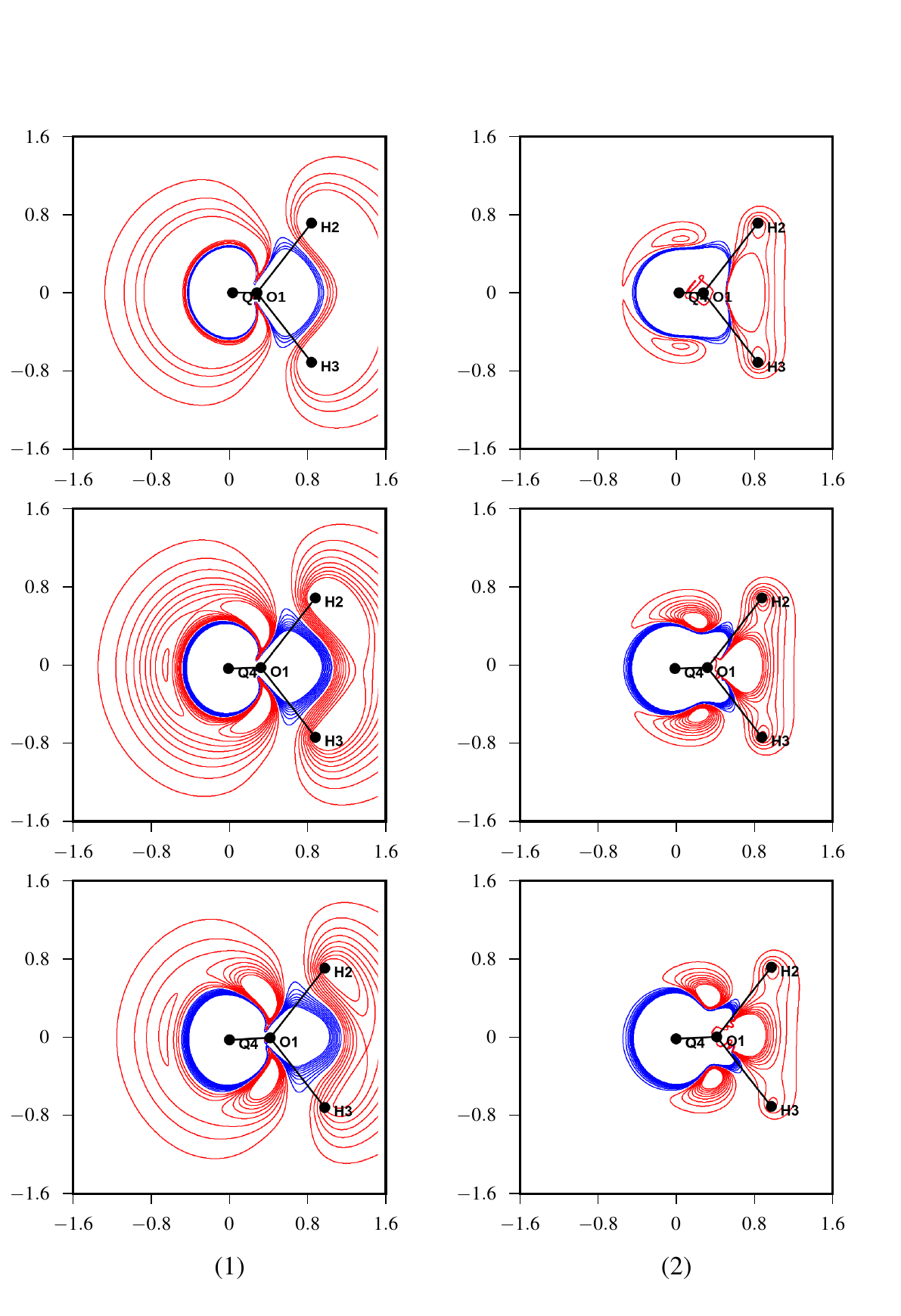}{0cm}{5mm}{0cm}{17mm}{20cm}{.9\linewidth}
\end{minipage}\qquad
\begin{minipage}[h]{.48\linewidth}
  \figA{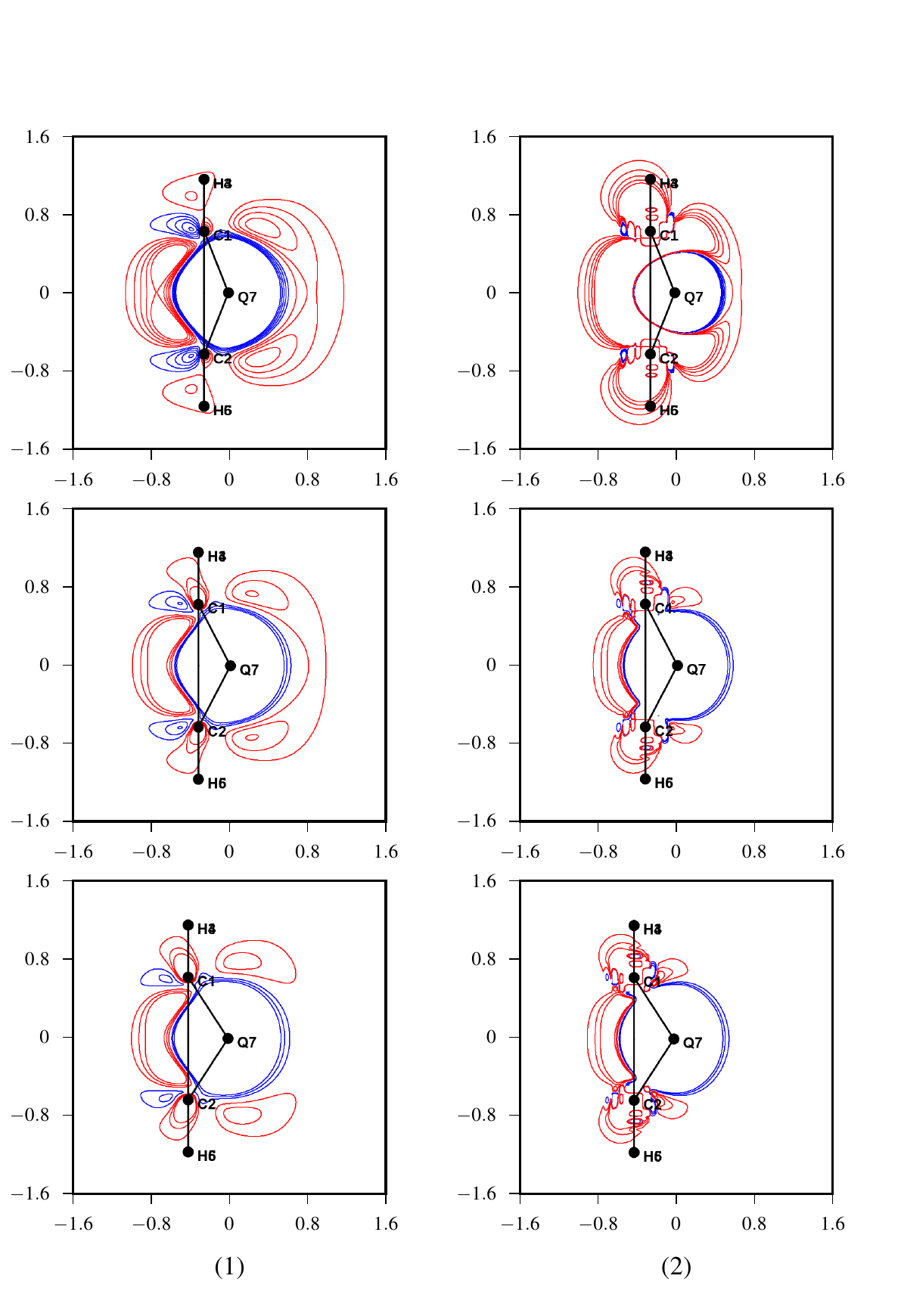}{0cm}{5mm}{0cm}{17mm}{20cm}{.9\linewidth}
\end{minipage}
\caption{Effect of the choice of the reference point, $\bf{r}_0$, denoted by Q4 
 and Q7, respectively for the water (left) and ethylene (right) molecules. 
 The reference point is moving
 outwards along a symmetry axis of the molecules, remaining in the 
 same bond domain. Column (1) represents the response function in the $xy$
 plane, column (2) illustrate the mathematical object defined by the rhs
 of \Ref{eq:chiApprox}, both shown for the reference point $\bf{r}_0$. }
\label{chiREF}
\end{figure*}

 The  comparison of the butadiene and the naphthalene clearly
 shows the difference in the extent of the delocalization in these 
 two systems. While in  the former case the 
 response function describes an induced change 
 of the charge which extends out only slightly from a central 
 perturbation to the terminal carbon atoms, in the case of the
 naphthalene  the charge density is  quite significantly perturbed
 on the entire molecular skeleton. We can observe also an  
 alternating sign of the charge density variations. The topology 
 of the corresponding independent particle form factor function 
 shows a reasonable resemblance with the behavior of the 
 linear response function, although some features 
 seem to be different on the two contour plots. For instance, the 
 multiple maxima observed in the response function plot of the naphthalene 
 molecule seems to be absent on the form factor plot. In the case of 
 the butadiene the strong positive peaks at the on the opposite 
 side of the molecular plane with respect to the reference point under 
 the C1 and C2 carbon atoms are almost invisible on the form factor plot.

 The contour plots in Figure  \FRef{chiALL} depend on the choice of the 
 reference point. We have examined the sensitivity of the response function 
 and of the form factor with respect to the position of the reference point. 
 Since the shape of these functions is expected to change radically if
 the reference point is moved from one electron pair domain to another, 
 we focused our attention to relatively small displacements within the
 same domain. As illustrated by Figure \FRef{chiREF}, the shape of the 
 response function does not change appreciably; we have the positive 
 peak, associated in the form factor model with the Dirac delta function
 and a negative region related  to the exchange hole in the form factor 
 model. However, according to the selected examples, the water
 and ethylene molecules, 
 some details may change in the case of a shift of the reference 
 point along the symmetry axis  
 and additional critical points may appear in the 
 negative (red) regions of both
 the response function and the form factor model.

%%%%%%%%%%%%%%%%%%%%%%%%%%%%%%%%%%%%%%%%%%%%%%%%%
\subsection{Exchange hole and localized orbitals}
%%%%%%%%%%%%%%%%%%%%%%%%%%%%%%%%%%%%%%%%%%%%%%%%%
Figure \FRef{hxALL} illustrates the relationship seen \Ref{eq:Hxloc3}
between the localized orbital and the exchange hole, by plotting
\begin{align}
h_x(\rA,\rR)
\quad \text{and} \quad
-\left| \phi_k(\rA) \right|^2
.
\nonumber
\end{align}

 The reference points  $\rR$ for the exchange hole have been  
 chosen at the centroids of the localized orbital $\phi_k$, obtained
 by the Foster-Boys localization criterium.
 The following orbitals were selected:
 the C-C bonds for the case of the ethylene molecule, 
 the C-C simple bond of the butadiene molecule,
 one of the oxygen lone pairs for the water molecule and 
 the central C-C bond of the naphthalene molecule.
 The grids are constructed such that the reference points 
 coincide with one of the grid points.
 
 In the case of the small water and ethylene molecules and 
 even in butadiene the contour plots of the exchange hole 
 and of the square of the  localized orbital are practically 
 indistinguishable. Although the agreement is very good in 
 naphthalene as well, the squared localized orbital seems to be 
 slightly more extended on the naphthalene rings than the 
 exchange hole.

 It has been observed for a long time that \textit{bond 
 polarizabilities} are remarkably transferable from one system 
 to another. This point was discussed from the theoretical 
 chemistry viewpoint by Claverie \cite{Claverie:78} by identifying 
 bond polarizabilities and localized orbital polarizabilities. The 
 isomorphism of the localized orbitals and of the exchange hole,
 which in turn is related to the local charge density response 
 puts this observation in a new context a provides a sort of 
 explanation in terms of the model independent exchange hole 
 through its close relationship with the linear response function
 itself.
 
 \subsection{Discussion} 
 
 From a conceptual point of view it is quite obvious that the notion of
 delocalization is primarily  connected to the question "How 
 electrons respond to an external disturbance?". The basic physical 
 observable in this respect is the charge density response function, 
 unfortunately its direct use in routine studies is made difficult by
 some obstacles. 
 
 The main difficulty is that the response function 
 depends on two space variables so its full
 pictorial representation in the three dimensional space or 
 in any lower dimensional cross-section (plane or line) 
 should be done  as a function of the reference point, which 
 should, in principle, scan the whole space. Geerlings and his 
 co-workers have recently published a series of articles, which 
 present and discuss response functions  (linear response kernels)
 for a series of systems \cite{Geerlings:14}. These studies have been 
 done most often on the noninteracting response function, which requires 
 significantly less computational resources, than the analogous 
 plots of the interacting
 response, e.g.\ at the coupled perturbed Kohn-Sham or Hartree-Fock 
 or correlated (coupled cluster) level.
 
 A way to get rid of the second variable would be to construct domain-averaged quantities,
 in analogy to the domain-averaged Fermi holes
 (DAFH) \cite{Ponec:99,Ponec:97a,Ponec:98a}:
 \begin{equation}
   \chi_{\Omega_A}(\ru) = \int_{\Omega_A} d\rd \,\chi(\ru,\rd).
 \end{equation}

 To the best of our knowledge this quantity constructed from
 the response function has not yet been proposed in the literature. 
 For instance, it is expected
 that the shape of the atomic domain-averaged response function
 provides information about space regions where the electrons of this
 particular atom are expected to be delocalized. A detailed study
 of such domain-averaged response functions will be be 
 the subject of a forthcoming publication.
 
 A further simplification in the characterization of the linear 
 response function can be achieved by a double domain average, 
 leading to a quantity identical to the 
 charge-flow polarizability in the distributed
 polarizability formalism:
 \begin{equation}
   \chi_{\Omega_A,\Omega_B} = \int_{\Omega_A} d\rd
    \int_{\Omega_B} d\ru \,\chi(\ru,\rd).
 \end{equation}

 $\chi_{\Omega_A,\Omega_B}$ measures the propensity of the charge 
 density in $\Omega_A$ to flow towards the domain $\Omega_B$ 
 under the effect of an electric potential difference between 
 the two domains. Application of the above definition to QTAIM domains
 defined by Bader \cite{Bader:book} leads to atom-atom charge flow 
 polarizabilities \cite{Angyan:94c,Haettig:96}. Although relatively small
 systems were studied in these works, clear signatures of strong 
 delocalization have been observed for the dicyan (NCCN) and benzene 
 molecules. In the former case, the end-to-end  charge flow is remarkably 
 high (-0.499) which is to be compared to the charge flow polarizability
 between bonded nitrogen and carbon atoms (-1.256) while the
 charge flow between non bonded carbon and nitrogen  atoms
 is very small (-0.042). In the benzene, the strong para charge-flow 
 is remarkable (-0.316) and the opposite sign meta charge flow (+0.103). 
 Such a behavior is in good agreement with well-known reactivity rules
 in substituted benzenes. 
 
 Our attempt to model the main features of the response function
 in terms of the density-weighted exchange hole was motivated 
 in a great extent by  the observation that the charge flow derived from
 the linear response function by a double atomic domain average 
 correlates well with de\-lo\-ca\-li\-za\-tion indices, 
 which can be calculated 
 from the exchange hole. Although the model presented in this work does
 correctly reflect the main trends, it can and should be improved in 
 the future. One possible way consists in finding a simple model for
 the position-dependent effective excitation energies. Work in this 
 direction is in progress.

 A recent exhaustive review by Geerlings and his coworkers
 \cite{Geerlings:14} came to our knowledge after completing our work.
 Their paper discusses in depth various chemical applications of the 
 linear response function, like mesomeric effects, delocalization, 
 aromaticity, reactivity, \etc of the response functions and 
 quantities derived therefrom.

 One of the roles of the density-weighted exchange hole is to 
 establish a link between the most general delocalization measures,
 the response function and the localized orbitals, which are 
 usually considered as outdated. The quality of the conventional localized 
 orbitals depends on the localization criteria. It is interesting to 
 notice that one can derive one-electron functions, which play an 
 analogous role as the localized orbitals, directly from the exchange
 hole. In this respect, one should mention the Fermi hole based 
 localized orbitals of Luken and Culberson \cite{Luken:84,Luken:90}, 
 the natural orbitals derived from the domain-averaged Fermi hole 
 by Pone{\u c} \cite{Ponec:98a}, the domain natural orbitals (DNO) 
 \cite{Francisco:09}, which were further generalized by diagonalization 
 of the $n$-th order cumulant density matrices, leading to Natural Adaptive
 Orbitals (NAdOs) \cite{Francisco:13}. In this respect one should mention 
 Cioslowski's isopycnic transformation method to obtain localized 
 natural orbitals from correlated (many-de\-ter\-mi\-nan\-tal) wave functions, 
 based on the invariance of the Fermi hole \cite{Cioslowski:90a}.

\section{Conclusions}
\label{sec:conclusions}

 The basic physical model that the (de)localization of 
 electrons is intimately related to fundamental physical 
 observables like the charge density response function
 of the system on the one hand and to the exchange-correlation 
 hole function on the other, has been illustrated on a few 
 selected examples. Note that in independent particle theories 
 we can consider the exchange hole and the noninteracting
 linear response function, which are much simpler to calculate 
 than the full linear response and the full many-body 
 exchange-correlation hole.
 
 Pictures of  two-dimensional cross-sections of the 
 molecular space provide a detailed insight to the relationships 
 which have been established earlier between charge-flow 
 polarizabilities and atom-atom delocalization indices. 
 Further and more rigorous comparisons should be done in 
 the future, since the  visual shape of the contour 
 plots may be strongly dependent on  the choice of the contours, 
 selected here on a linear scale and uniformly for all the 
 presented pictures. The still objectively existing discrepancies 
 are expected to be removed by an improved, position-dependent 
 effective excitation energy modulating the form factor function, 
 as indicated previously.  

 The relationship between localized orbitals and the exchange hole,
 first observed by Luken more than 30 years ago, permits an 
 \textit{a posteriori}
 justification of the use of localized orbitals
 in qualitative interpretations of bonding and localization.
 In the mean time this analysis underlines some inherent 
 limitations of the localized orbital picture. 

 There are several physical implications of the relationships
 which we attempted to make more plausible via graphical
 illustrations.
 First, we can see that the charge density of an 
 electronic system at point $\rd$ responding to an external 
 perturbation applied in  $\ru$ can be quite reasonably 
 predicted from the exchange hole. Roughly speaking, the response 
 will be nonzero essentially in those region where the 
 exchange hole is nonzero too. 
 Furthermore, it may be surprising that the
 charge density response in genuinely localized systems remains 
 local in real space: it extends only to a few-atom region around 
 the perturbation. An implication of this observation is that 
 the charge-flow between such regions, which are in close 
 resemblance with the exchange hole and by consequence to  
 localized orbital domains, can be relatively small. Selecting such
 domains for a multi-center, distributed description of 
 intermolecular forces, in  particular in the case of induction 
 and dispersion
 interactions, leads naturally to models without significant charge-flow
 contributions. In contrast, in metal-like strongly delocalized systems the
 possible necessity of including charge-flow contributions
 \cite{Misquitta:10} becomes
 obvious by the failure of finding localized orbitals or in 
 more general terms,  by the inherently delocalized nature of the 
 density-weighted exchange hole. Since the Resta localization tensor
 \cite{Resta:06b} or using a recently suggested alternative  
 name \cite{Brea:13}, the Total Position Spread Tensor (TPST), is 
 the second moment of the density-weighted exchange hole, one can 
 establish also a direct link \cite{Resta:11} to the well-known
 near-sightedness concept of Walter Kohn \cite{Kohn:64,Prodan:05}, 
 which provides a general framework to discuss localization and 
 delocalization in solids but also in finite molecular systems.



 \section*{Acknowledgement}
 J.G.A. thanks for the support of this research by the 
 European Union and the State of Hungary, co-financed by the 
 European Social Fund in the framework of TÁMOP 4.2.4. 
 A/2-11-1-2012-0001 ‘National Excellence Program’.

%%\bibliographystyle{elsarticle-num} 
%%\bibliography{MusAng-2015}


\end{document}